\documentclass[twocolumn,showpacs,preprintnumbers,amsmath,amssymb,prl]{revtex4-1}

\usepackage{graphicx}
\usepackage{xcolor, colortbl}
\usepackage{pdfpages}
\usepackage{hyperref}
\usepackage{hhline}

\hypersetup{colorlinks=true, linkcolor=blue, citecolor=red, urlcolor=magenta, pdfauthor={D. M. Juraschek}}

\makeatletter
    \def\CT@@do@color{%
      \global\let\CT@do@color\relax
            \@tempdima\wd\z@
            \advance\@tempdima\@tempdimb
            \advance\@tempdima\@tempdimc
    \advance\@tempdimb\tabcolsep
    \advance\@tempdimc\tabcolsep
    \advance\@tempdima2\tabcolsep
            \kern-\@tempdimb
            \leaders\vrule
                    \hskip\@tempdima\@plus  1fill
            \kern-\@tempdimc
            \hskip-\wd\z@ \@plus -1.05fill }
\makeatother

\definecolor{IronBlue}{rgb}{0.33,0.49,0.88}

\definecolor{ErbiumRed}{rgb}{0.55,0.22,1}

\definecolor{OxygenRed}{rgb}{1,0.24,0.1}

\definecolor{OctahedronYellow}{rgb}{0.84,0.84,0.1}
\newcommand{\octahedronyellow}{OctahedronYellow!15!white}

\newcommand{\reff}[1]{(\ref{#1})}
\newcommand{\fig}[1]{\text{Fig.~\ref{#1}}}
\newcommand{\tab}[1]{\text{Tab.~\ref{#1}}}

\newcommand{\erfeo}{\ensuremath{\text{ErFeO}_3}}
\newcommand{\units}[2]{\ensuremath{#1\,\text{#2}}}
\newcommand{\Ag}{\ensuremath{\text{A}_{\text{g}}}}
\newcommand{\Aeinsg}{\ensuremath{\text{A}_{\text{1g}}}}
\newcommand{\Au}{\ensuremath{\text{A}_{\text{u}}}}
\newcommand{\Bxu}[1]{\ensuremath{\text{B}_{#1\text{u}}}}
\newcommand{\Bxg}[1]{\ensuremath{\text{B}_{#1\text{g}}}}
\newcommand{\Eu}{\ensuremath{\text{E}_{\text{u}}}}
\newcommand{\QIR}[1][\phantom{2}]{\ensuremath{Q_{\text{IR}}^{#1}}}
\newcommand{\QR}[1][\phantom{2}]{\ensuremath{Q_{\text{R}}^{#1}}}
\newcommand{\Qeins}[1][\phantom{2}]{\ensuremath{Q_{\text{IR}_1}^{#1}}}
\newcommand{\Qzwei}[1][\phantom{2}]{\ensuremath{Q_{\text{IR}_2}^{#1}}}
\newcommand{\wR}[1][]{\ensuremath{\omega_{\text{R}}^{#1}}}
\newcommand{\weins}[1][]{\ensuremath{\omega_{1}^{#1}}}
\newcommand{\wzwei}[1][]{\ensuremath{\omega_{2}^{#1}}}
\newcommand{\ce}[1]{\ensuremath{c_{\text{#1}}}}
\newcommand{\de}[1]{\ensuremath{d_{\text{#1}}}}

\newcommand{\ang}{\ensuremath{\text{\AA}}}
\newcommand{\fromto}{--}
\newcommand{\vecc}[1]{\ensuremath{\textbf{#1}}}
\newcommand{\diff}[1]{\ensuremath{\text{d}#1}}
\newcommand{\expp}[1]{\ensuremath{\text{e}_{\text{\phantom{X}}}^{#1}}}
\renewcommand{\deg}{\ensuremath{^{\circ}}}

\begin{document}

\title{Ultrafast Structure Switching through Nonlinear Phononics}
\author{D.~M.\ Juraschek}
\email{dominik.juraschek@mat.ethz.ch}
\affiliation{Materials Theory, ETH Zurich, CH-8093 Z\"{u}rich, Switzerland }
\author{M.\ Fechner}
\affiliation{Materials Theory, ETH Zurich, CH-8093 Z\"{u}rich, Switzerland }
\author{N.~A.\ Spaldin}
\affiliation{Materials Theory, ETH Zurich, CH-8093 Z\"{u}rich, Switzerland }

\begin{abstract}
We describe an ultrafast coherent control of the transient structural distortion arising from nonlinear phononics in \erfeo{}. 
Using density functional theory, we calculate the structural properties as input to an anharmonic phonon model that describes 
the response of the system to a pulsed optical excitation. We find that the trilinear coupling of two orthogonal infrared-active 
phonons to a Raman-active phonon causes a transient distortion of the lattice. The direction of the distortion is determined 
by the polarization of the exciting light, suggesting a route to nonlinear phononic lattice control and switching. Since the 
occurrence of the coupling is determined by the symmetry of the system we propose that it is a universal feature of orthorhombic 
and tetragonal perovskites.
\end{abstract}

\maketitle


Over the last decade it has been shown repeatedly that laser excitation of infrared-active phonons is a powerful tool for 
modifying the properties of materials. This {\it dynamical materials design} approach has been used to drive metal-insulator 
transitions \cite{rini:2007, tobey:2008}, to melt orbital order \cite{beaud:2009, caviglia:2012} and to induce superconductivity 
or modify superconducting transition temperatures \cite{fausti:2011, hu:2014} in a range of complex oxides. Particularly 
intriguing is the case in which the laser intensity is so high that the usual harmonic approximation for the lattice dynamics 
breaks down and anharmonic phonon-phonon interactions become important. Recent experimental and theoretical studies 
\cite{forst:2011, mankowsky:2014} have clarified that quadratic-linear cubic coupling of the form \QIR[2]\QR{} between a driven 
infrared-active mode, \QIR{}, and a Raman-active mode, \QR{}, causes a shift in the equilibrium structure to a nonzero value 
of the Raman mode normal coordinates. This nonlinear phononic effect has most notably been associated with the observation of 
coherent transport, an indicator of superconductivity, far above the usual superconducting Curie temperature in underdoped 
YBaCu$_3$O$_{6+\delta}$ \cite{forst:2013, mankowsky:2015, fechner:2016}. 

Here we investigate theoretically a different kind of cubic phononic coupling of the trilinear form \Qeins\Qzwei\QR{}, in 
which two different infrared-active (IR) modes are excited simultaneously and couple anharmonically to a single Raman mode. 
Our motivation is provided by recent experimental work on the perovskite-structure orthoferrite \erfeo{} \cite{nova:2015}, 
in which two polar modes of similar frequencies with atomic displacement patterns along the inequivalent $a$ and $b$ orthorhombic 
axes were simultaneously excited. Ref.~\cite{nova:2015} reported and analyzed the resulting excitation of a magnon; here our 
focus is on the changes caused by and the implications of the nonlinear phonon dynamics.


\begin{figure}[t]
\includegraphics[scale=0.25]{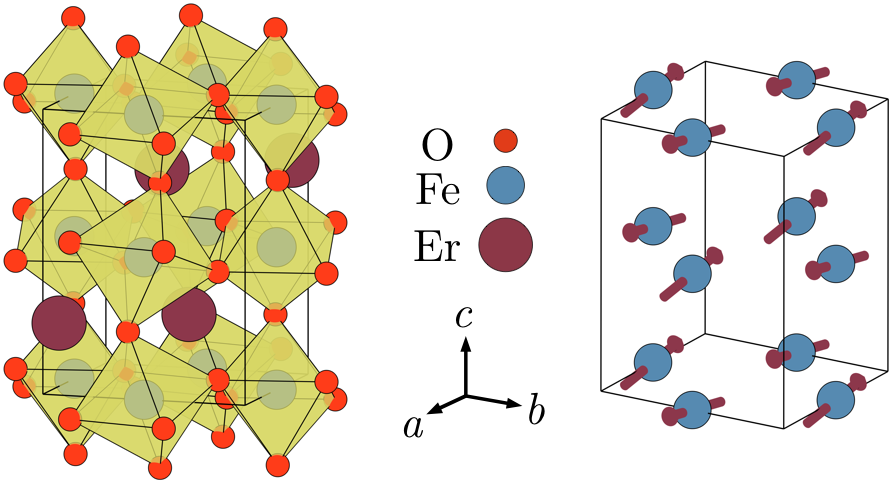}
\vspace{-5pt}
\caption{\label{fig:Pnma}$Pnma$ lattice structure of \erfeo{}. The left unit cell shows the tilted octahedra of the distorted 
perovskite structure. The right unit cell shows the G-type antiferromagnetic ordering of the iron spins along the $a$ axis.}
\vspace{-10pt}
\end{figure}

\erfeo{} is a distorted perovskite with the orthorhombic $Pnma$ structure and the typical G-type antiferromagnetic ordering 
of the Fe$^{3+}$ magnetic moments \cite{treves:1965} (\fig{fig:Pnma}). The primitive magnetic unit cell contains 20 atoms, 
resulting in 60 phonon modes characterised by representations (within the orthorhombic point group $mmm$) \Ag{}, \Bxg{(1,2,3)}, 
\Au{} and \Bxu{(1,2,3)}. Of the polar ``u'' modes, only \Bxu{1}, \Bxu{2} and \Bxu{3} have dipole moments and are therefore 
excitable by mid-infrared light. 

The totally symmetric representation in the $mmm$ point group is \Ag{}, and so coupling is only symmetry allowed between 
combinations of phonon modes whose product contains the \Ag{} representation. To cubic order, and for excitation of the \Bxu{} 
modes, this is the case for two types of mode combinations: quadratic-linear B$_{i\text{u}}^{2}$A$_{\text{g}}^{\phantom{i}}$ 
and trilinear \Bxu{i}\Bxu{j}\Bxg{k}, where $\{i,j,k\}=\{1,2,3\}, i\neq{}j\neq{}k$. The quadratic-linear case describes the 
coupling of a single optically excited polar mode to a symmetric \Ag{} Raman mode, and is the nonlinear phononic scenario that 
has been studied previously \cite{mankowsky:2014, forst:2013, subedi:2014, mankowsky:2015, subedi:2015, fechner:2016}. In this 
work we focus instead on the trilinear case in which two IR modes with different symmetries are excited and combine with a 
\Bxg{k} mode. We note also that in quartic order, any combination of infrared and Raman modes of the biquadratic form 
\QIR[2]\QR[2] results in an \Ag{} representation, and so the potential energy up to fourth order can be written as:
\begin{align}\label{eq:potential}
V(\vecc{Q})          = & \frac{\weins[2]}{2}\Qeins + \frac{\wzwei[2]}{2}\Qzwei + \frac{\wR[2]}{2}\QR + \nonumber \\
                       & \ce{R}\QR[3] + \ce{1R}\Qeins[2]\QR + \ce{2R}\Qzwei[2]\QR + \nonumber \\
                       & \ce{12R}\Qeins\Qzwei\QR + \\
                       & \frac{\de{1}}{4}\Qeins[4] + \frac{\de{2}}{4}\Qzwei[4] + \frac{\de{R}}{4}\QR[4] + \nonumber \\
                       & \de{12}\Qeins[2]\Qzwei[2] + \de{1R}\Qeins[2]\QR[2] + \de{2R}\Qzwei[2]\QR[2], \nonumber
\end{align}
where \Qeins{} and \Qzwei{} denote the amplitudes of IR modes with different symmetries and eigenfrequencies \weins{} and 
\wzwei{} and \QR{} is the amplitude of a Raman mode with eigenfrequency \wR{}. The coefficients $\{c\}$ and $\{d\}$ define 
the strengths of the cubic and quartic anharmonicity and are material specific. As stated above, $\ce{R}=\ce{1R}=\ce{2R}=0$ 
by symmetry if \QR{} corresponds to a \Bxg{i} mode and $\ce{12R}=0$ if \QR{} corresponds to an \Ag{} mode. 

We will see that the quartic coefficients in ErFeO$_3$ are small, consistent with earlier work for related transition metal 
oxides \cite{fechner:2016}, and since we are interested in isolating the effects of the trilinear coupling, we analyze the 
reduced potential 
\begin{align}
V(\vecc{Q}) = &\frac{\weins[2]}{2}\Qeins + \frac{\wzwei[2]}{2}\Qzwei + \frac{\wR[2]}{2}\QR + \nonumber \\
              &\ce{12R}\Qeins\Qzwei\QR
\end{align}
in the following. The value of \QR{} that minimizes $V(\vecc{Q})$, and which corresponds to the average structure induced 
by the trilinear coupling, is obtained trivially from this expression as 
\begin{equation}\label{eq:trilineardistortion}
Q_{\text{R}_\text{min}}^{\phantom{2}} = \frac{\ce{12R}\Qeins\Qzwei}{\wR[2]}.
\end{equation}
We therefore expect that the induced structural distortions will be largest for low-frequency Raman modes with large 
\ce{12R} coupling coefficients. 


We begin by calculating the structural properties of ErFeO$_3$ from first-principles within the density functional formalism
as implemented in the Vienna ab-initio simulation package (VASP) \cite{kresse:1996, kresse2:1996}. We used the default VASP 
PAW pseudopotentials with valence electronic configurations Er\,(6$s^2$5$p^6$5$d^1$), Fe\,(3$d^7$4$s^1$) and O\,(2$s^2$2$p^4$), 
with the $4f$ electrons of erbium in the core. Treatment of the $4f$ electrons as core states has the desirable side effect 
of yielding the room-temperature magnetic structure, with the iron spins oriented along the $a$ axis and a weak ferromagnetic 
moment along $c$ \cite{treves:1965} (\fig{fig:Pnma}), in our zero kelvin calculation, since the experimentally observed 
low-temperature spin-reorientation transitions to other easy axes \cite{white:1969}, attributed to interaction with the Er 
$4f$ moments, are suppressed. Good convergence was obtained with a plane-wave energy cut-off of \units{850}{eV} and a 
6$\times$6$\times$4 $k$-point mesh to sample the Brillouin zone. We converged the Hellmann-Feynman forces to 
\units{10^{-5}}{eV/\ang{}} for the calculation of phonons with the frozen-phonon method as implemented in the phonopy package 
\cite{phonopy}. For the exchange-correlation functional we chose the PBEsol \cite{PBEsol} form of the generalized gradient 
approximation (GGA) with a Hubbard $U$ correction on the Fe $3d$ states. We found that an on-site Coulomb interaction of 
$U=\units{3.7}{eV}$ and a Hund's exchange of $J=\units{0.7}{eV}$ optimally reproduce both the lattice dynamical properties 
\cite{subbarao:1970, koshizuka:1980, nova:2015} and the G-type antiferromagnetic ordering \cite{treves:1965} as well as the 
photoemission spectrum of closely related LaFeO$_3$ \cite{wadati:2005}. In particular we found that phonon eigenfrequencies 
are underestimated by other approaches, including the usual PBE functionals. Our fully relaxed structure with lattice constants 
$a=\units{5.19}{\ang}$, $b=\units{5.56}{\ang}$ and $c=\units{7.52}{\ang}$ fits reasonably well to the experimental values of 
Ref.~\cite{eibschutz:1965}, as do our calculated phonon eigenfrequencies. Anharmonic coupling constants were computed by 
calculating the total energies as a function of ion displacements along the normal mode coordinates of every \QR{} mode and
of every \Qeins{} and \Qzwei{} modes that it couples to and then fitting the resulting three-dimensional energy landscape to 
the potential $V$ of Eqn.~\reff{eq:potential}. 


\begin{table}[t]
\caption{Calculated and experimental phonon eigenfrequencies in THz. Infrared data were taken from 
Refs.~\cite{subbarao:1970, nova:2015}, Raman data from Ref.~\cite{koshizuka:1980}.}
\begin{tabular}{lrr|lrr}
\hline\hline
Sym. & ~~DFT~~ & ~~Exp.~~ & Sym. & ~~DFT~~ & ~~Exp.~~ \\
\hline
\rowcolor{\octahedronyellow}
\Ag{} & 3.3 & 3.4 & \Bxg{1} & 3.2 & 3.4 \\
    & 4.0 & 4.0 &         & 4.8 & 4.8 \\
\rowcolor{\octahedronyellow}
    & 8.1 & 8.1 &         & 9.6 & 9.7 \\
    & 10.0 & 10.0 &         & 10.5 & -- \\
\rowcolor{\octahedronyellow}
    & 12.5 & 12.4 &         & 14.6 & 15.1 \\
    & 13.0 & 13.0 &         & 16.2 & -- \\
\rowcolor{\octahedronyellow}
    & 14.9 & 14.9 &         & 18.3 & -- \\
\hline
\rowcolor{\octahedronyellow}
\Bxu{2} & 3.1 & -- & \Bxu{3} & 3.5 & -- \\
        & 5.7 & -- &         & 5.2 & -- \\
\rowcolor{\octahedronyellow}
        & 7.2 & -- &         & 7.5 & -- \\
        & 8.9 & -- &         & 8.6 & -- \\
\rowcolor{\octahedronyellow}
        & 9.7 & -- &         & 9.9 & -- \\
        & 10.2 & -- &         & 10.9 & 10.9 \\
\rowcolor{\octahedronyellow}
        & 13.2 & 13.3 &         & 12.4 & -- \\
        & 15.7 & -- &         & 15.5 & -- \\
\rowcolor{\octahedronyellow}
        & 16.0 & 16.2 &         & 16.5 & 17.0 \\
\hline\hline
\end{tabular}
\label{tab:frequencies}
\vspace{-10pt}
\end{table}

\begin{figure}[t]
\includegraphics[scale=0.22]{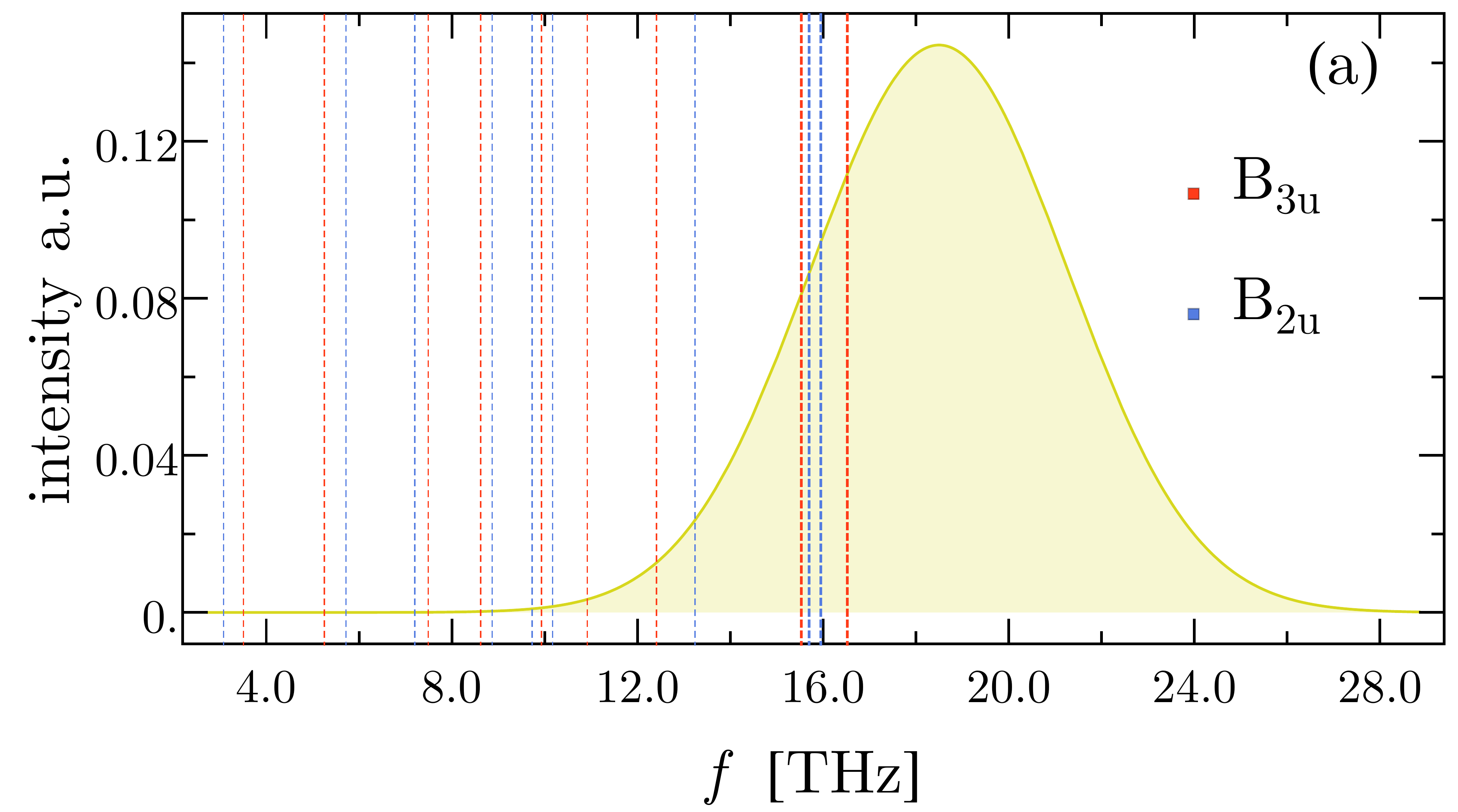}\phantom{X}\\
\includegraphics[scale=0.26]{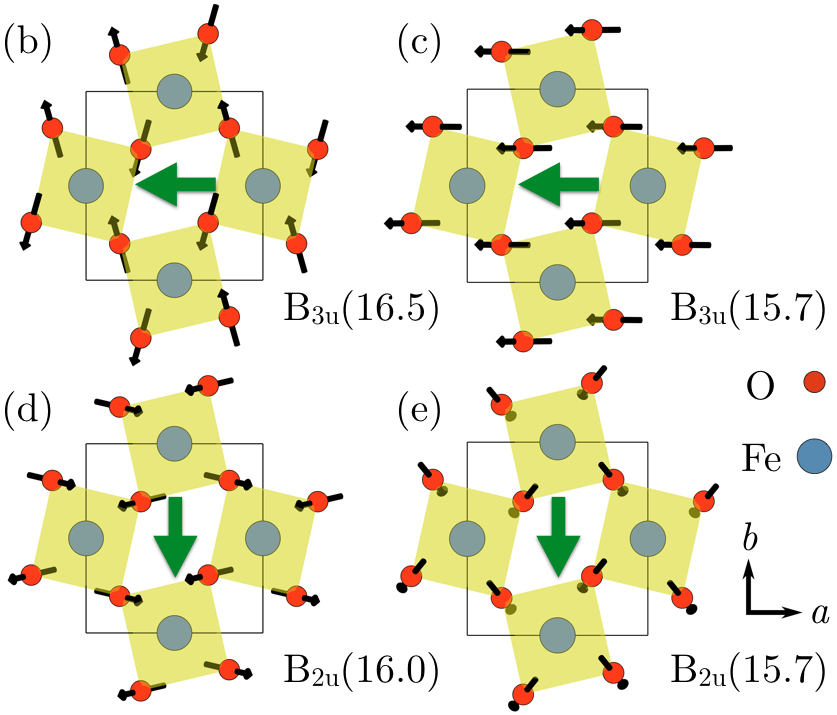}
\vspace{-5pt}
\caption{\label{fig:pumpspectrum}\textbf{(a)} Frequency spectrum of a model pump pulse with mean frequency 
$\omega_0=\units{18.5}{THz}$ and a full width at half maximum of \units{6.5}{THz} as used in Ref.~\cite{nova:2015}. 
Our calculated phonon eigenfrequencies for \erfeo{} are shown as vertical dashed lines, with the four highest frequency 
modes that we consider in this work marked as thicker dashed lines. The right panels show the displacements of the oxygen 
ions in modes \textbf{(b)} \Bxu{3}(16.5), \textbf{(c)} \Bxu{2}(16.0), \textbf{(d)} \Bxu{2}(15.7) and \textbf{(e)} 
\Bxu{3}(15.5) as thin black arrows and the resulting direction of polarization as thick green arrows.}
\vspace{-10pt}
\end{figure}

In the experiment of Ref.~\cite{nova:2015}, the laser pulse was directed perpendicular to the short axes $a$ and $b$ of the 
orthorhombic \erfeo{} crystal. The phonons that are excited by such a pulse have symmetries \Bxu{3} (polarization along the 
$a$ axis) and \Bxu{2} (polarization along the $b$ axis). The trilinear coupling term is therefore only nonzero for Raman 
modes of \Bxg{1} symmetry; both IR modes couple quadratic-linearly to \Ag{} Raman modes. The experimental pulse frequency 
was $\omega_0=\units{18.5}{THz}$ with a pulse width of $\sigma_\omega=\units{2.8}{THz}$, so that IR modes between around 15 
and 20 THz are significantly excited by the pulse. Our calculated values for the phonon eigenfrequencies with symmetries 
\Bxu{3}, \Bxu{2}, \Ag{} and \Bxg{1} are listed in \tab{tab:frequencies} (for the full list of calculated eigenfrequencies 
see the Appendix) along with available experimental values. In \fig{fig:pumpspectrum} \textbf{a} we show a model mid-infrared 
laser pulse with the properties of that used in Ref.~\cite{nova:2015} and indicate our calculated \Bxu{3} and \Bxu{2} 
eigenfrequencies with vertical lines. We see that phonon modes with \Bxu{3} and \Bxu{2} symmetries occur in pairs of similar 
eigenfrequencies, consistent with the small orthorhombicity of \erfeo{} (in a tetragonal structure they would form a pair of 
degenerate \Eu{} modes). It is also clear from \fig{fig:pumpspectrum} \textbf{a} that only the group of four IR modes with 
the highest eigenfrequencies \Bxu{3}(16.5), \Bxu{2}(16.0), \Bxu{2}(15.7) and \Bxu{3}(15.5) are significantly excited by the 
pulse of Ref.~\cite{nova:2015}. We show the displacements of the oxygen ions in the eigenvectors of these modes, and the 
direction of the corresponding polarization in \fig{fig:pumpspectrum} \textbf{b}\fromto\textbf{e}. In the following analysis 
we focus on the two highest frequency modes, \Bxu{3}(16.5) and \Bxu{2}(16.0) and the lowest frequency Raman modes, \Bxg{1}(3.2) 
and \Ag{}(3.3), for which we expect the biggest effect according to Eqn.~\reff{eq:trilineardistortion}. 

Our calculated values for the anharmonic coefficients $\{c\}$ and $\{d\}$ for these modes are shown in \tab{tab:coupling} 
with the list of coefficients for the remaining combinations of the four highest frequency IR modes given in the Appendix. 
We see that the quartic order coupling coefficients between Raman and IR modes, \de{1R} and \de{2R} are all at least one 
order of magnitude smaller than the cubic coupling coefficients, \ce{1R}, \ce{2R} and \ce{12R}. (Note that the other 
anharmonic coefficents listed for completeness do not couple Raman and IR modes). This confirms our expectation that 
phonon-phonon coupling of the biquadratic kind is negligible for the dynamics of this system. We see also that the 
coefficient of trilinear coupling to the \Bxg{1} mode is similar in magnitude (in fact slightly larger) to that of the 
quadratic-linear coupling to the \Ag{}  mode. 

\begin{table}[b]
\vspace{-11pt}
\caption{Anharmonic coefficients for the two Raman modes with lowest eigenfrequencies that couple to the \Bxu{3}(16.5) 
and \Bxu{2}(16.0) IR modes. Units are meV/$(\ang\sqrt{\mu})^{n}$, where $\mu$ is the atomic mass unit and $n$ the order 
of the phonon amplitude. The $\{c\}$ coefficients with 0 value vanish due to symmetry arguments, whereas \de{1R} for 
the \Bxg{1} mode is accidentally zero.}
\begin{tabular}{|l|r|r|r|r|r|r|r|r|r|r|}
\hline
& \multicolumn{1}{c|}{\ce{R}} 
& \multicolumn{1}{c|}{\ce{1R}} 
& \multicolumn{1}{c|}{\ce{2R}} 
& \multicolumn{1}{c|}{\ce{12R}} 
& \multicolumn{1}{c|}{\de{1}} 
& \multicolumn{1}{c|}{\de{2}} 
& \multicolumn{1}{c|}{\de{R}} 
& \multicolumn{1}{c|}{\de{12}} 
& \multicolumn{1}{c|}{\de{1R}} 
& \multicolumn{1}{c|}{\de{2R}}\\
\hline
\Bxg{1}(3.2) 
& \multicolumn{1}{c|}{0}
& \multicolumn{1}{c|}{0}
& \multicolumn{1}{c|}{0}
& \cellcolor{IronBlue!80!white}$-$10.2 
& \cellcolor{OctahedronYellow!80!red}17.6 
& \cellcolor{OctahedronYellow!80!white}8.2 
& \cellcolor{OctahedronYellow!20!white}1.1 
& \cellcolor{OctahedronYellow!80!white}12.0 
& \cellcolor{OctahedronYellow!20!white}0.0 
& \cellcolor{IronBlue!20!white}$-$0.1\\
\hline
\Ag{} (3.3) 
& \cellcolor{IronBlue!20!white}$-$0.5 
& \cellcolor{OctahedronYellow!80!white}7.8 
& \cellcolor{OctahedronYellow!80!white}3.7 
& \multicolumn{1}{c|}{0}
& \cellcolor{OctahedronYellow!80!red}17.6 
& \cellcolor{OctahedronYellow!80!white}8.2 
& \cellcolor{OctahedronYellow!20!white}1.1 
& \cellcolor{OctahedronYellow!80!white}12.0 
& \cellcolor{IronBlue!20!white}$-$0.3 
& \cellcolor{OctahedronYellow!20!white}0.1\\
\hline
\end{tabular}\\
\vspace{4pt}
\begin{tabular}{c|c|c|c|c|c|c|c|c}
\hhline{~|*7-}
$-20$ 
& \cellcolor{IronBlue!80!black} \phantom{H} 
& \cellcolor{IronBlue!80!white} \phantom{H} 
& \cellcolor{IronBlue!20!white} \phantom{H} 
& \cellcolor{white} \phantom{H} 
& \cellcolor{OctahedronYellow!20!white} \phantom{H} 
& \cellcolor{OctahedronYellow!80!white} \phantom{H} 
& \cellcolor{OctahedronYellow!80!red} \phantom{H} 
& $20$ \\
\cline{2-8}
\end{tabular}
\label{tab:coupling}
\end{table}

\begin{figure*}[t]
\begin{minipage}{0.74\textwidth}
\includegraphics[scale=0.21]{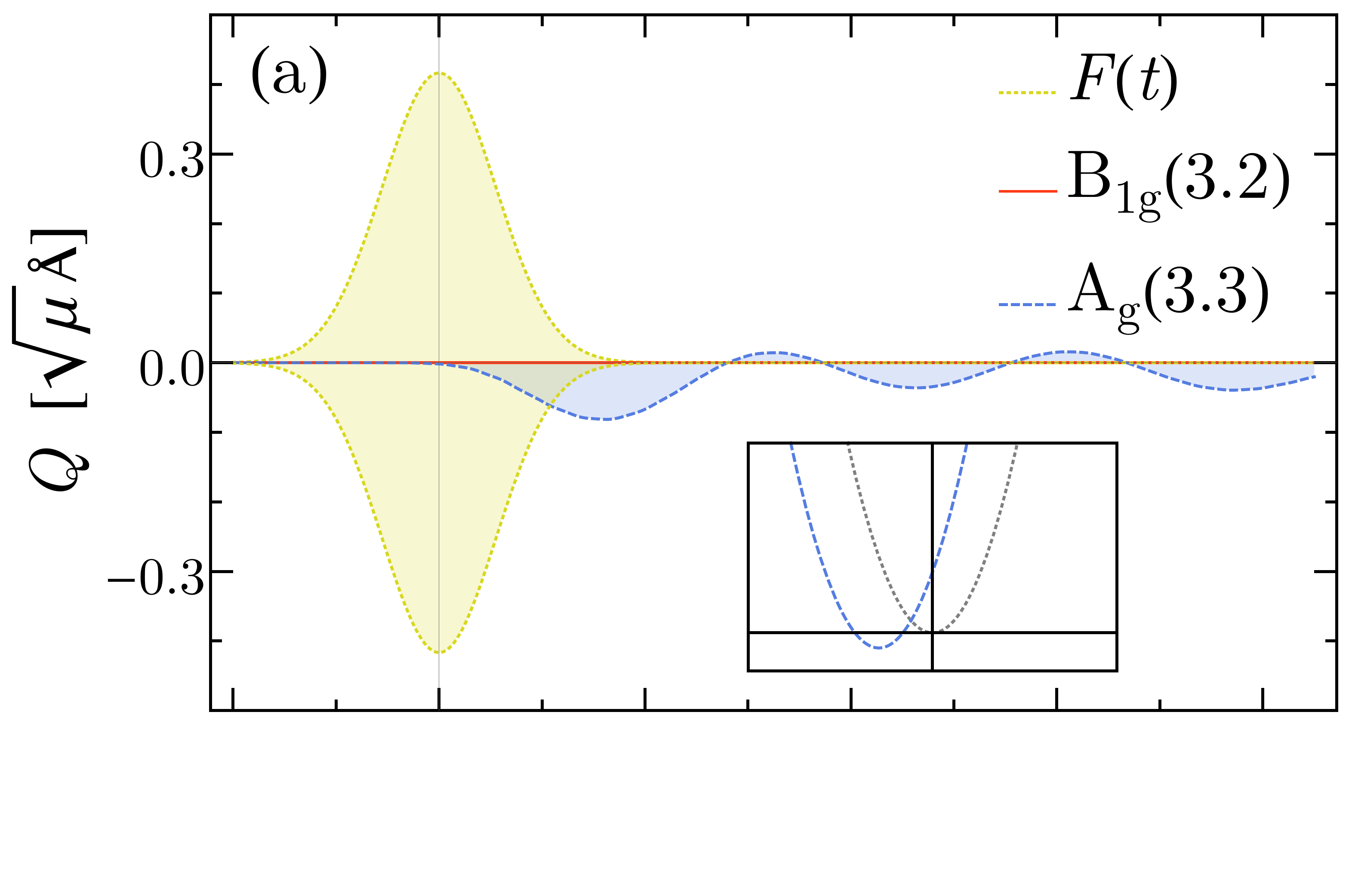}
\hspace{-5pt}
\includegraphics[scale=0.21]{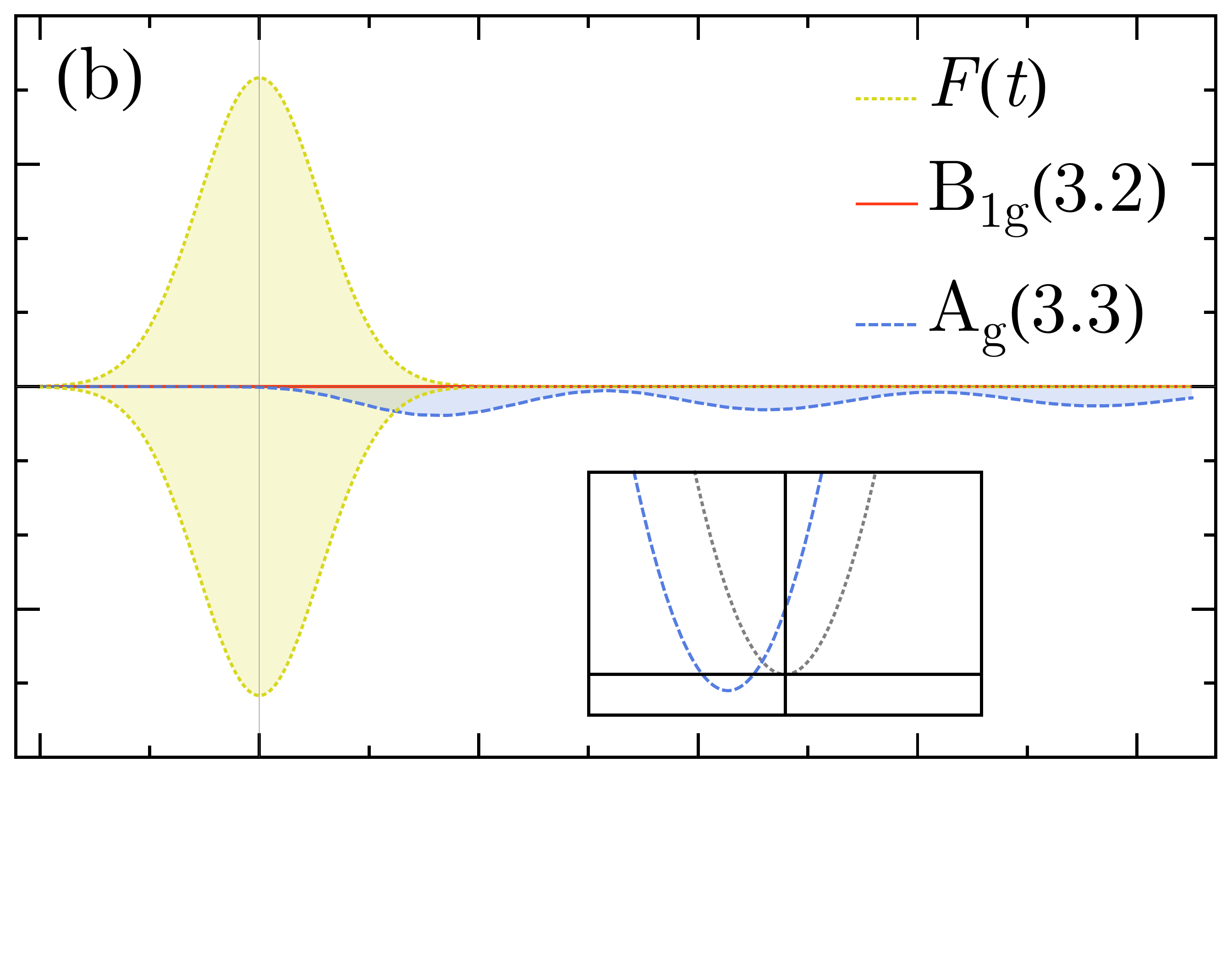}\\
\vspace{-25pt}
\includegraphics[scale=0.21]{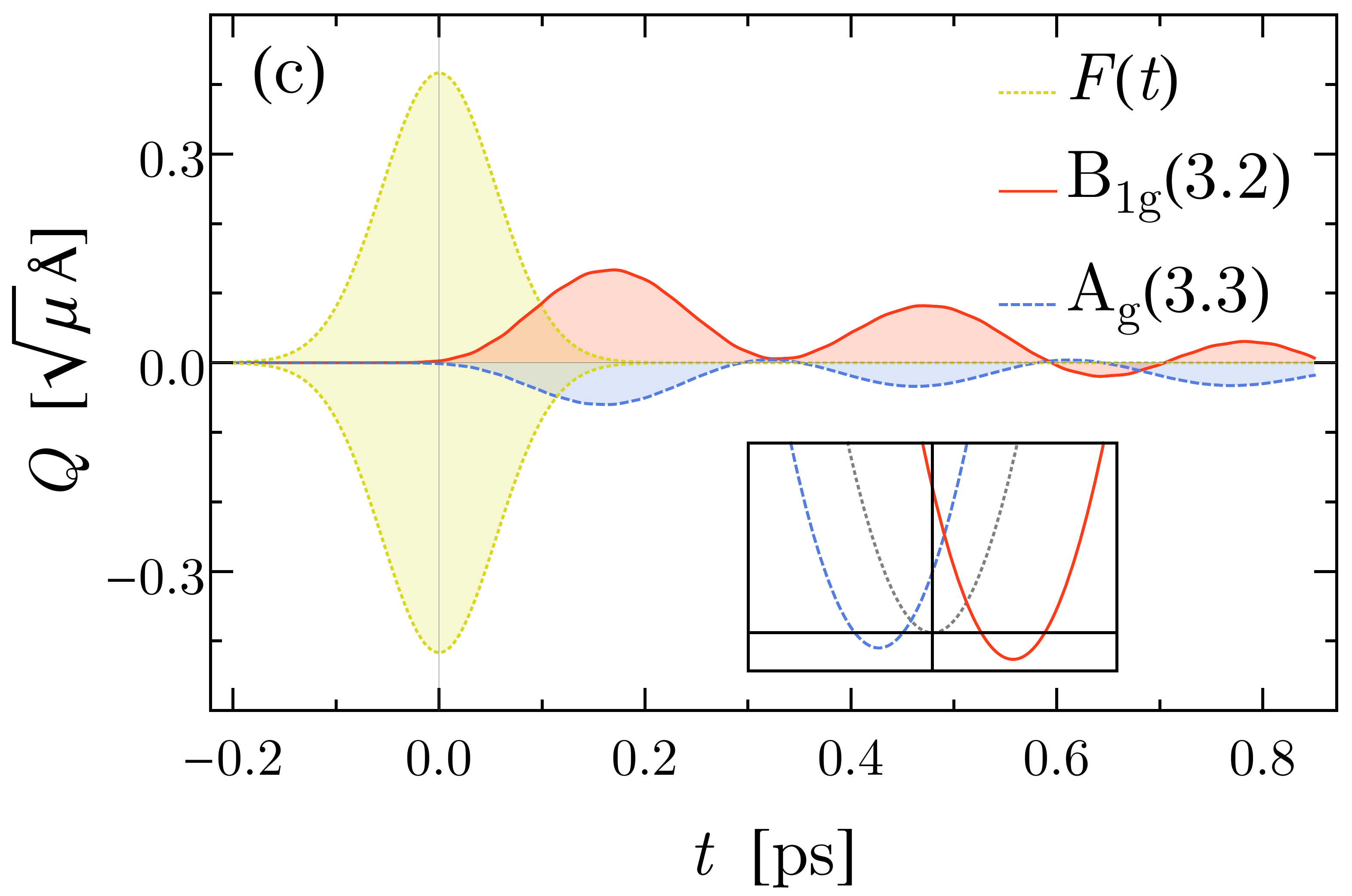}
\hspace{-5pt}
\includegraphics[scale=0.21]{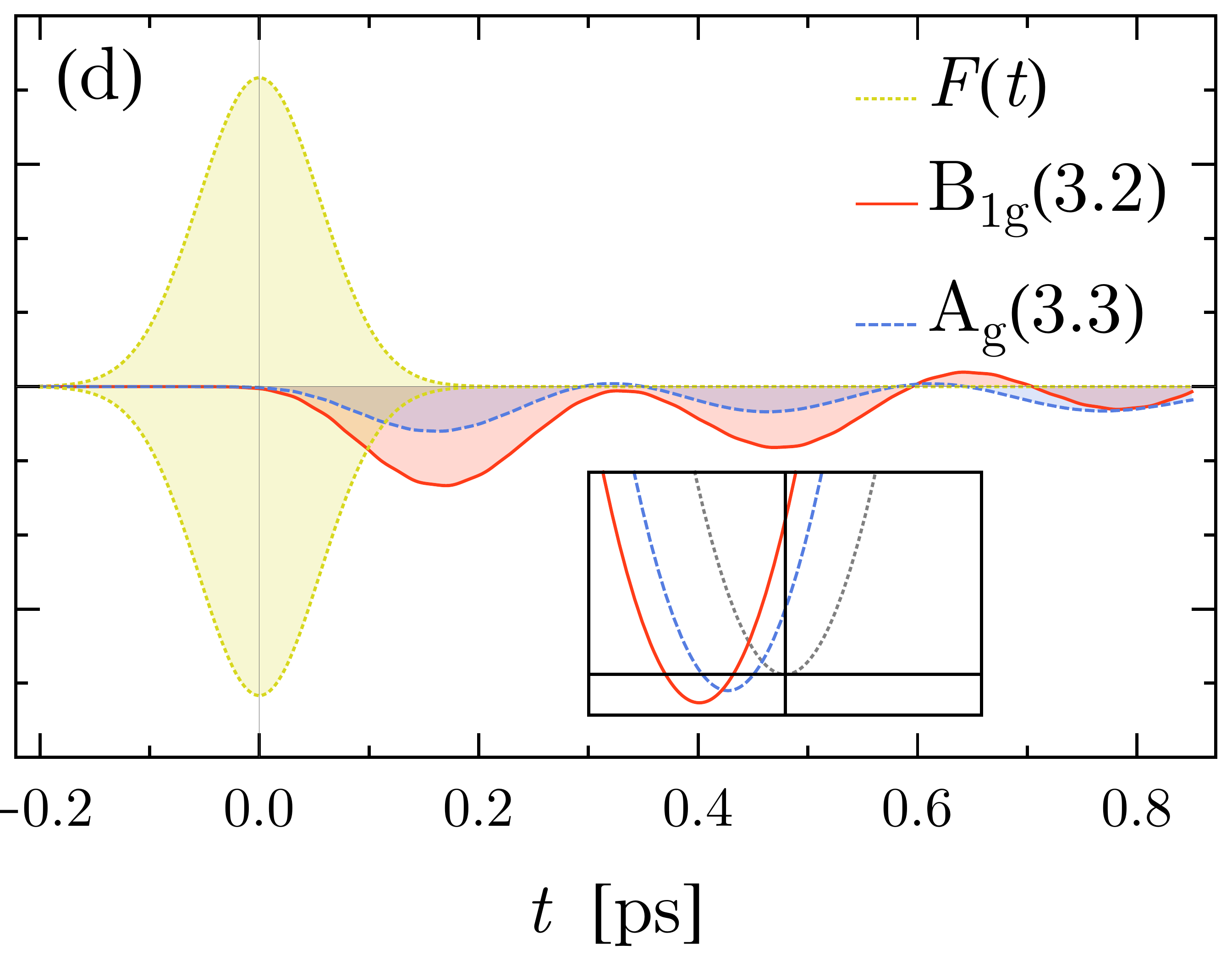}
\end{minipage}
\begin{minipage}{0.25\textwidth}
\includegraphics[scale=0.26]{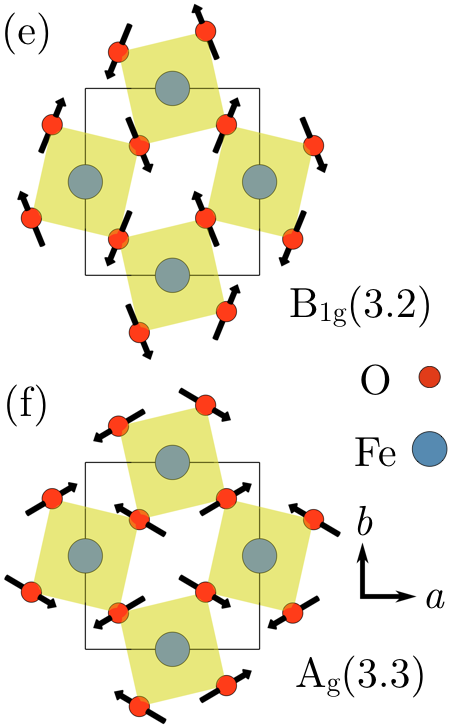}
\vspace{17pt}
\end{minipage}
\vspace{-5pt}
\caption{\label{fig:evolution}Evolution of the two Raman modes with lowest eigenfrequencies when pumped with 
linearly-polarized light at \textbf{(a)} $\theta=0\deg$, \textbf{(b)} $\theta=90\deg$, \textbf{(c)} $\theta=+45\deg$ 
and \textbf{(d)} $\theta=-45\deg$, with $\theta=0\deg$ corresponding to a polarization along the $a$ axis of the 
crystal and $\theta=90\deg$ to a polarization along the $b$ axis. We assumed a realistic linewidth of 
$\gamma\approx{}f/20$ \cite{nova:2015}. The shifts in the minima of the Raman modes are shown schematically in the 
insets, with the gray curve indicating an unshifted potential, and blue and red the shifted \Ag{} and \Bxg{1} potential 
respectively. The right panel shows the displacements of the oxygen ions corresponding to the modes \textbf{(e)} 
\Bxg{1}(3.2) and \textbf{(f)} \Ag{}(3.3).}
\vspace{-10pt}
\end{figure*}

To investigate the evolution of the anharmonic system, we next solve numerically the dynamical equations of motion that 
form the system of coupled differential equations:
\begin{equation}
\ddot{\vecc{Q}} + \gamma\dot{\vecc{Q}} + \nabla_{\vecc{Q}} \left[ V(\vecc{Q}) - F(t,\theta)\QIR{} \right] = 0,
\end{equation}
where $\vecc{Q}=(\Qeins{},\Qzwei{},\QR{})$ describes both the excited IR modes and one coupled Raman mode. $\gamma$ 
is the linewidth (inverse lifetime) of each mode and $F(t)$ the driving force on the IR modes that represents the 
laser pulse. We model the laser pulse with both time and frequency broadening as
\begin{equation}
F(t,\theta{}) = F_{0} h(t,t_{0}) \cos(\theta) \int\limits_{-\infty}^{\infty} \diff{\omega} \sin(\omega{}t) h(\omega{},\omega_{0}),
\end{equation}
where $F_{0}$ is the maximum intensity and $h(x,x_{0})=\expp{-(x-x_0)^2/(2\sigma_{x}^{2})}/(\sqrt{2\pi}\sigma_{x})$ 
provides a gaussian spread both in time ($x=t$) and frequency ($x=\omega$). To conform to the experiment of Ref.~\cite{nova:2015} 
we set the pulse duration to $\sigma_{t}=\units{55}{fs}$ and the frequency broadening to $\sigma_\omega=\units{2.8}{THz}$ 
at a mean frequency of $\omega_0=\units{18.5}{THz}$ with a peak amplitude of $F_0=\units{10}{MV/cm}$. $\theta$ is the 
polarization angle of the linearly-polarized light from the laser with respect to the symmetry of the IR modes with 
$\theta=0\deg$ corresponding to a polarization along the $a$ axis of the crystal and $\theta=90\deg$ to a polarization 
along the $b$ axis. The evolution of the system after an excitation with the laser pulse is shown in \fig{fig:evolution} 
\textbf{a}\fromto\textbf{d} for a range of polarization angles. The displacements of the oxygen ions corresponding to the 
\Bxg{1}(3.2) and \Ag{}(3.3) Raman modes are shown in \fig{fig:evolution} \textbf{e}, \textbf{f}.

In \fig{fig:evolution} \textbf{a} and \textbf{b} the polarization of the light pulse is along one of the lattice vectors 
so that in each case only IR modes of one symmetry type are excited, \Bxu{3} in \textbf{a}, where the pulse is oriented 
along the $a$ axis ($\theta=0\deg$) and \Bxu{2} in \textbf{b}, for a pulse along the $b$ axis ($\theta=90\deg$). As a 
result there is no trilinear coupling and no excitation of the \Bxg{1}(3.2) mode. In both cases, the \Ag{}(3.3) mode is 
excited through its quadratic-linear coupling to the single IR mode. We see that its sinusoidal oscillation (blue line) 
is not centered around zero amplitude indicating the characteristic transient structural distortion caused by the 
quadratic-linear coupling \QIR[2]\QR{}, as observed previously in 
Refs.~\cite{forst:2013, mankowsky:2015, subedi:2014, subedi:2015, fechner:2016} and described above. The induced shifts 
of the minimum in the potential for the Raman mode are shown in the insets. Since the potential depends quadratically on 
the IR mode, and the signs of the coupling coefficients \ce{1R} and \ce{2R} are the same, the direction of the structural 
distortion is independent of the angle of its polarization, with the same direction of shift for $\theta=0\deg$ and 
$\theta=90\deg$. Note, however, that the strength of the quadratic-linear coupling as different in the two cases, since 
the values of the coupling coefficients \ce{1R} and \ce{2R} differ.

In \fig{fig:evolution} \textbf{c} and \textbf{d} the polarization of the light pulse is midway between the lattice vectors,
at $\theta=+45\deg$ and $\theta=-45\deg$ respectively so that both IR modes are excited simultaneously. In this case the 
response of the \Bxg{1} mode is maximal. The behavior of the \Ag{} mode is the same as the previous cases, with the 
amplitude and direction of the shift in the average value independent of the polarization of the pulse. The trilinear 
coupling \Qeins\Qzwei\QR{} of the \Bxg{1}(3.2) mode shows strikingly a different behaviour, however. We see that when the 
polarization angle is changed from $\theta=+45\deg$ to $\theta=-45\deg$, the trilinear coupling term changes sign and the 
transient deformation of the lattice is in the opposite direction along the the normal mode coordinates of the \Bxg{1} mode.

The time for which the transient structural deformation of the \Bxg{1} mode maintains its initial direction is determined 
by the inverse difference frequency $|\weins-\wzwei|^{-1}$, which determines the time-scale of the dephasing. The smaller 
the difference frequency the IR modes, the longer it takes them to dephase and thus the longer the directional selectivity 
of the trilinear coupling persists. For the realistic linewidth that we assumed, the structure relaxes back to the ground 
state before the IR modes dephase noticeably. In the case of a tetragonal structure in which the in-plane IR modes form a 
degenerate \Eu{} pair, $\weins=\wzwei$ and no dephasing occurs. In this limit both quadratic-linear \textit{and} trilinear 
coupling to a fully symmetric \Aeinsg{} Raman mode should occur simultaneously, with their relative strengths determined 
by the angle of the excitation pulse to the crystallographic axes.


In summary, we have shown that excitation of two infrared-active (IR) phonon modes with different symmetries but similar 
eigenfrequencies in the orthorhombic perovskite ErFeO$_3$ causes a transient structural distortion along the eigenvectors 
of a coupled \Bxg{1} Raman mode as a result of its trilinear coupling with the IR modes. In contrast to the quadratic-linear 
coupling of a symmetric \Ag{} Raman mode to a {\it single} IR mode that has been discussed previously 
\cite{mankowsky:2014, forst:2013, subedi:2014, mankowsky:2015, subedi:2015, fechner:2016} and which we also observe here, 
the direction of the transient distortion is determined by the polarization of the excitation pulse relative to the 
crystallographic axes and can be reversed by reversing the polarization direction. While the analysis presented here was 
performed for ErFeO$_3$, it is directly applicable to all orthorhombic and tetragonal perovskites, with the strengths of 
the coupling constants and the values of the phonon frequencies of course being material dependent; extension to other 
crystal classes involves a further straightforward symmetry analysis. Our results suggest that nonlinear phononics can be 
used to control and switch the orientation of induced transient crystal structures. 

We thank T.~Nova and A.~Cavalleri for fruitful discussions. This work was supported by the ETH Z\"urich and by the ERC 
Advanced Grant program, No. 291151. Calculations were performed at the Swiss National Supercomputing Centre (CSCS) 
supported by the project IDs s624 and p504.

\bibliography{literature}

\clearpage

\appendix*

\section*{Appendix}

In \tab{tab:fullcoupling} we show the full list of anharmonic coefficients used in this work. Note that the values may vary 
slightly for different mappings of the energy landscape due to the finite grid size. For example \de{1} for the \Bxu{3}(16.5) 
mode is \units{17.6}{meV/$(\ang\sqrt{\mu})^{n}$} when calculated together with \Bxu{2}(16.0) and \Bxg{1}(3.2), but 
\units{17.2}{meV/$(\ang\sqrt{\mu})^{n}$} together with \Bxu{2}(16.0) and \Bxg{1}(3.2). For the dynamical equations of motion 
we used the average of these values. In \tab{tab:fulleigenfrequencies} we show the full list of calculated eigenfrequencies 
in units of terahertz and inverse centimetres. 

\begin{table}[b]
\vspace{-15pt}
\caption{Full list of anharmonic coefficients for the two Raman modes with lowest eigenfrequencies that couple to the four infrared 
modes with highest eigenfrequencies. Units are meV/$(\ang\sqrt{\mu})^{n}$, where $\mu$ is the atomic mass unit and $n$ the order of 
the phonon amplitude.}
\begin{tabular}{|l|r|r|r|r|r|r|r|r|r|r|}
\hline
& \multicolumn{1}{c|}{\ce{R}} 
& \multicolumn{1}{c|}{\ce{1R}} 
& \multicolumn{1}{c|}{\ce{2R}} 
& \multicolumn{1}{c|}{\ce{12R}} 
& \multicolumn{1}{c|}{\de{1}} 
& \multicolumn{1}{c|}{\de{2}} 
& \multicolumn{1}{c|}{\de{R}} 
& \multicolumn{1}{c|}{\de{12}} 
& \multicolumn{1}{c|}{\de{1R}} 
& \multicolumn{1}{c|}{\de{2R}}\\
\hline\hline
\multicolumn{11}{|c|}{\Qeins=\Bxu{3}(16.5), \Qzwei=\Bxu{2}(16.0)} \\
\hline
\Bxg{1}(3.2) 
& \multicolumn{1}{c|}{0}
& \multicolumn{1}{c|}{0}
& \multicolumn{1}{c|}{0}
& \cellcolor{IronBlue!80!white}$-$10.2 
& \cellcolor{OctahedronYellow!80!red}17.6 
& \cellcolor{OctahedronYellow!80!white}8.2 
& \cellcolor{OctahedronYellow!20!white}1.1 
& \cellcolor{OctahedronYellow!80!white}12.0 
& \cellcolor{OctahedronYellow!20!white}0.0 
& \cellcolor{IronBlue!20!white}$-$0.1\\
\hline
\Ag{} (3.3) 
& \cellcolor{IronBlue!20!white}$-$0.5 
& \cellcolor{OctahedronYellow!80!white}7.8 
& \cellcolor{OctahedronYellow!80!white}3.7 
& \multicolumn{1}{c|}{0}
& \cellcolor{OctahedronYellow!80!red}17.6 
& \cellcolor{OctahedronYellow!80!white}8.2 
& \cellcolor{OctahedronYellow!20!white}1.1 
& \cellcolor{OctahedronYellow!80!white}12.0 
& \cellcolor{IronBlue!20!white}$-$0.3 
& \cellcolor{OctahedronYellow!20!white}0.1\\
\hline\hline
\multicolumn{11}{|c|}{\Qeins=\Bxu{3}(16.5), \Qzwei=\Bxu{2}(15.7)} \\
\hline
\Bxg{1}(3.2) 
& \multicolumn{1}{c|}{0}
& \multicolumn{1}{c|}{0}
& \multicolumn{1}{c|}{0}
& \cellcolor{OctahedronYellow!80!white}6.5 
& \cellcolor{OctahedronYellow!80!red}17.2 
& \cellcolor{OctahedronYellow!80!white}3.1 
& \cellcolor{OctahedronYellow!20!white}0.6 
& \cellcolor{OctahedronYellow!80!white}8.4 
& \cellcolor{OctahedronYellow!20!white}0.1 
& \cellcolor{OctahedronYellow!20!white}0.1\\
\hline
\Ag{} (3.3) 
& \cellcolor{IronBlue!20!white}$-$0.4 
& \cellcolor{OctahedronYellow!80!white}7.6 
& \cellcolor{OctahedronYellow!80!white}5.9 
& \multicolumn{1}{c|}{0}
& \cellcolor{OctahedronYellow!80!red}17.3 
& \cellcolor{OctahedronYellow!80!white}3.1 
& \cellcolor{OctahedronYellow!20!white}0.6 
& \cellcolor{OctahedronYellow!80!white}8.3 
& \cellcolor{IronBlue!20!white}$-$0.2 
& \cellcolor{IronBlue!20!white}$-$0.1\\
\hline\hline
\multicolumn{11}{|c|}{\Qeins=\Bxu{3}(15.5), \Qzwei=\Bxu{2}(16.0)} \\
\hline
\Bxg{1}(3.2) 
& \multicolumn{1}{c|}{0}
& \multicolumn{1}{c|}{0}
& \multicolumn{1}{c|}{0}
& \cellcolor{IronBlue!80!black}$-$18.0 
& \cellcolor{OctahedronYellow!80!red}22.9 
& \cellcolor{OctahedronYellow!80!white}6.8 
& \cellcolor{OctahedronYellow!20!white}0.2 
& \cellcolor{OctahedronYellow!80!white}8.5 
& \cellcolor{OctahedronYellow!20!white}0.8 
& \cellcolor{OctahedronYellow!20!white}0.1\\
\hline
\Ag{} (3.3) 
& \cellcolor{IronBlue!20!white}$-$0.6 
& \cellcolor{OctahedronYellow!20!white}0.8 
& \cellcolor{OctahedronYellow!80!white}3.8 
& \multicolumn{1}{c|}{0}
& \cellcolor{OctahedronYellow!80!red}23.0 
& \cellcolor{OctahedronYellow!80!white}6.9 
& \cellcolor{OctahedronYellow!20!white}0.4 
& \cellcolor{OctahedronYellow!80!white}8.4 
& \cellcolor{OctahedronYellow!20!white}0.4 
& \cellcolor{OctahedronYellow!20!white}0.2\\
\hline\hline
\multicolumn{11}{|c|}{\Qeins=\Bxu{3}(15.5), \Qzwei=\Bxu{2}(15.7)} \\
\hline
\Bxg{1}(3.2) 
& \multicolumn{1}{c|}{0}
& \multicolumn{1}{c|}{0}
& \multicolumn{1}{c|}{0}
& \cellcolor{IronBlue!80!black}$-$22.0 
& \cellcolor{OctahedronYellow!80!red}22.8 
& \cellcolor{OctahedronYellow!20!white}2.3 
& \cellcolor{OctahedronYellow!20!white}0.0 
& \cellcolor{OctahedronYellow!80!white}8.5 
& \cellcolor{OctahedronYellow!20!white}0.8 
& \cellcolor{OctahedronYellow!20!white}0.3\\
\hline
\Ag{} (3.3) 
& \cellcolor{IronBlue!20!white}$-$0.3 
& \cellcolor{OctahedronYellow!20!white}0.3 
& \cellcolor{OctahedronYellow!80!white}5.6 
& \multicolumn{1}{c|}{0}
& \cellcolor{OctahedronYellow!80!red}22.8 
& \cellcolor{OctahedronYellow!20!white}2.3 
& \cellcolor{OctahedronYellow!20!white}0.1 
& \cellcolor{OctahedronYellow!80!white}8.5 
& \cellcolor{OctahedronYellow!20!white}0.4 
& \cellcolor{OctahedronYellow!20!white}0.0\\
\hline
\end{tabular}
\begin{tabular}{c|c|c|c|c|c|c|c|c}
\hhline{~|*7-}
$-25$ 
& \cellcolor{IronBlue!80!black} \phantom{H} 
& \cellcolor{IronBlue!80!white} \phantom{H} 
& \cellcolor{IronBlue!20!white} \phantom{H} 
& \cellcolor{white} \phantom{H} 
& \cellcolor{OctahedronYellow!20!white} \phantom{H} 
& \cellcolor{OctahedronYellow!80!white} \phantom{H} 
& \cellcolor{OctahedronYellow!80!red} \phantom{H} 
& $25$ \\
\cline{2-8}
\end{tabular}
\label{tab:fullcoupling}
\end{table}

\pagebreak

\begin{table}[!h]
\vspace{-15pt}
\caption{Full list of calculated phonon frequencies.}
\begin{tabular}{lrrr||lrrr}
\hline\hline
\# & ~THz & ~cm$^{-1}$ & ~~Sym.~~ & \# & ~THz & ~cm$^{-1}$ & ~~Sym.~~ \\
\hline\hline
\rowcolor{\octahedronyellow}
   1  &     0  &      0  &  acoust.  &  31  &     9.7  &     324  &  \Bxu{2}  \\
   2  &     0  &      0  &  acoust.  &  32  &     9.9  &     331  &  \Bxu{3}  \\
   \rowcolor{\octahedronyellow}
   3  &     0  &      0  &  acoust.  &  33  &    10.0  &     332  &  \Ag{}  \\
   4  &   2.2  &     75  &  \Au{}    &  34  &    10.2  &     339  &  \Bxu{2}  \\
   \rowcolor{\octahedronyellow}
   5  &   3.1  &    103  &  \Bxu{2}  &  35  &    10.5  &     349  &  \Bxg{1}  \\
   6  &   3.2  &    107  &  \Bxg{1}  &  36  &    10.8  &     361  &  \Bxu{1}  \\
   \rowcolor{\octahedronyellow}
   7  &   3.3  &    111  &  \Ag{}    &  37  &    10.9  &     364  &  \Bxu{3}  \\
   8  &   3.5  &    117  &  \Bxu{3}  &  38  &    10.9  &     365  &  \Bxg{3}  \\
   \rowcolor{\octahedronyellow}
   9  &   3.6  &    119  &  \Bxg{2}  &  39  &    11.1  &     370  &  \Au{}  \\
  10  &   3.9  &    129  &  \Bxg{3}  &  40  &    12.4  &     414  &  \Bxu{3}  \\
  \rowcolor{\octahedronyellow}
  11  &   4.0  &    134  &  \Ag{}    &  41  &    12.5  &     416  &  \Ag{}  \\
  12  &   4.7  &    155  &  \Bxu{1}  &  42  &    12.8  &     427  &  \Bxg{3}  \\
  \rowcolor{\octahedronyellow}
  13  &   4.7  &    158  &  \Au{}    &  43  &    12.9  &     432  &  \Bxg{2}  \\
  14  &   4.8  &    160  &  \Bxg{1}  &  44  &    13.0  &     432  &  \Ag{}  \\
  \rowcolor{\octahedronyellow}
  15  &   4.8  &    161  &  \Bxu{1}  &  45  &    13.2  &     442  &  \Bxu{2}  \\
  16  &   5.2  &    175  &  \Bxu{3}  &  46  &    13.9  &     465  &  \Bxg{2}  \\
  \rowcolor{\octahedronyellow}
  17  &   5.7  &    191  &  \Bxu{2}  &  47  &    14.5  &     484  &  \Au{}  \\
  18  &   5.9  &    196  &  \Au{}    &  48  &    14.6  &     487  &  \Bxg{1}  \\
  \rowcolor{\octahedronyellow}
  19  &   7.0  &    232  &  \Au{}    &  49  &    14.8  &     494  &  \Bxu{1}  \\
  20  &   7.2  &    240  &  \Bxu{2}  &  50  &    14.9  &     496  &  \Ag{}  \\
  \rowcolor{\octahedronyellow}
  21  &   7.5  &    250  &  \Bxu{3}  &  51  &    15.5  &     518  &  \Bxu{3}  \\
  22  &   7.7  &    257  &  \Bxu{1}  &  52  &    15.7  &     524  &  \Bxu{2}  \\
  \rowcolor{\octahedronyellow}
  23  &   7.7  &    257  &  \Bxg{3}  &  53  &    15.8  &     525  &  \Au{}  \\
  24  &   8.1  &    269  &  \Ag{}    &  54  &    15.8  &     527  &  \Bxu{1}  \\
  \rowcolor{\octahedronyellow}
  25  &   8.6  &    288  &  \Bxu{3}  &  55  &    16.0  &     532  &  \Bxu{2}  \\
  26  &   8.9  &    296  &  \Bxu{2}  &  56  &    16.2  &     540  &  \Bxg{1}  \\
  \rowcolor{\octahedronyellow}
  27  &   9.2  &    306  &  \Au{}    &  57  &    16.5  &     551  &  \Bxu{3}  \\
  28  &   9.3  &    309  &  \Bxu{1}  &  58  &    18.3  &     612  &  \Bxg{1}  \\
  \rowcolor{\octahedronyellow}
  29  &   9.4  &    313  &  \Bxg{2}  &  59  &    18.4  &     614  &  \Bxg{3}  \\
  30  &   9.6  &    320  &  \Bxg{1}  &  60  &    19.3  &     645  &  \Bxg{2}  \\
\hline\hline
\end{tabular}
\label{tab:fulleigenfrequencies}
\end{table}

\end{document}